\documentclass{aastex}
\usepackage{spr-astr-addons}

\RequirePackage{color}

\begin{document}

\title{The Mid-term and Long-term Solar Quasi-periodic Cycles and the Possible Relationship with Planetary Motions}
\slugcomment{Not to appear in Nonlearned J., 45.}
\shorttitle{Solar Cycles and Planetary Motions}

\shortauthors{Tan \& Cheng.}

\author{Baolin Tan\altaffilmark{1}} \and \author{Zhuo Cheng\altaffilmark{2}}
\altaffiltext{1}{Key Laboratory of Solar Activity, National
Astronomical Observatories, Chinese Academy of Sciences, Beijing
100012, China. Email: bltan@nao.cas.cn.} \altaffiltext{2}{Purple
Mountain Observatory, Chinese Academy of Sciences, Nanjing 210008,
China.}

\begin{abstract}
This work investigates the solar quasi-periodic cycles with
multi-timescales and the possible relationships with planetary
motions. The solar cycles are derived from long-term observations
of the relative sunspot number and microwave emission at frequency
of 2.80 GHz. A series of solar quasi-periodic cycles with
multi-timescales are registered. These cycles can be classified
into 3 classes: (1) strong PLC (PLC is defined as the solar cycle
with a period very close to the ones of some planetary motions,
named as planetary-like cycle) which is related strongly with
planetary motions, including 9 periodic modes with relatively
short period ($P<12$ yr), and related to the motions of the inner
planets and of Jupiter; (2) weak PLC, which is related weakly to
planetary motions, including 2 periodic modes with relatively long
period ($P>12$ yr), and possibly related to the motions of outer
planets; (3) non-PLC, which so far has no obvious evidence to show
the relationship with any planetary motions. Among planets,
Jupiter plays a key role in most periodic modes by its sidereal
motion or spring tidal motions with other planets. Among planetary
motions, the spring tidal motion of the inner planets and of
Jupiter dominates the formation of most PLCs. The relationships
between multi-timescale solar periodic modes and the planetary
motions will help us to understand the essential natures and
prediction of solar activities.
\end{abstract}


\section{Introduction}

Generally, the Sun is known to be the unique star dominating the
geo-space environment. It is very important to understand and
predict when and how the solar activity will take place in the
near future. The investigation of periodicity of solar activity is
a fundamental work for solar prediction. For this reason, we need
to find out and confirm each periodic mode of solar activity, the
correlations between the solar periodic activity and motions of
the celestial bodies. The latter includes solar internal motions,
external planetary motions, and the coupling among them. Such
investigation will help us to understand the real generation
mechanisms of solar activities. However, as we know, because of
lack of enough reliable long-term observations, uncertainties do
exist in almost all the deduced results in some extent, including
the solar quasi-periodic modes with various timescales. There are
a lot of works on the analysis of solar periodic modes, and their
generation mechanisms. And the latter is also including the
planetary influences (Wolff \& Patrone, 2010). In previous works,
many periodic modes are registered from the analysis of different
solar proxies (e.g., yearly averaged sunspot number, cosmogenic
isotopes, historic records, radiocarbon records in tree-rings,
etc.), for example, the 11 yr solar cycle, 51.5 yr period (Tan,
2011), 53 yr period (Le \& Wang, 2003), 80-90 yr period
(Gleissberg, 1971), the 65-130 yr quasi-periodic secular cycle
(Nagovitsyn, 1997), about 100 yr periodic cycle (Frick et al,
1997; Le \& Wang, 2003; Tan, 2011), 160-270 yr double century
cycle (Schove, 1979), 203 yr Suess cycle (Suess, 1980). Otaola \&
Zenteno (1983) proposed that long term cycles within the range of
80-100 and 170-180 yr are existed certainly.

The above solar periodic modes belong to long-term periodicity.
However, there are also some mid-term periodicity, which is
defined as the timescale between 27 d and 11 yr, the lower limit
means the period of solar rotation, and the upper limit is just
the period of the most famous solar Schwabe cycle. Numerous works
also reported the solar mid-term periodic modes. For example, 53
d, 85 d, 152 d, 248 d, 334 d, 683 d, etc. derived from the
analysis of solar flare index (Kilcik et al. 2010, etc.)

In this work, we make a detailed investigation of the
quasi-periodic modes with mid-term and long-term timescales of
solar activity by analysis of long-term data set of the microwave
emission at frequency of 2.80 GHz and the relative sunspot number.
The generation mechanisms of different solar periodic modes have
not yet been understood completely. We try to make a comprehensive
comparison between the solar periodic modes and the planetary
motions, and attempt to find out their relationships. The
planetary motions include the planetary sidereal motion and the
conjunction motions of planets. Section 2 introduces the
observation data and the corresponding analysis methods. Section 3
presents the main results of the periodicity analysis of solar
activity. Section 4 is the physical discussions of the
relationships between solar quasi-periodic modes and the planetary
orbital periods or conjunction motion of some planets. Finally,
conclusions are obtained in section 5.

\section{Data and Analysis Method}

\subsection{Data}

Generally, the relative sunspot number (RSN) is regarded as the
most important indicator of solar activities. It actually reflects
the intensity of solar magnetic activity, and the later always
dominates all the solar eruptions, such as solar flares, CMEs, and
filament eruptions, etc. Since 1700, an unbroken record of RSN is
accumulated. The data set of RSN can be downloaded from the
Solar-Geophysical Data (SGD) prompt reports at web site:
http://www.ngdc.noaa.gov. When we investigate the solar long-term
periodic activity, we may analyze the annual averaged RSN (ASN)
dataset during 1700-2011 with data length of $L=311$ yr. And when
we investigate the solar mid-term or short-term periodic activity
we may adopt the daily RSN records conveniently.

From the SGD prompt reports, we can also obtain the daily solar
total flux of microwave emission at frequency of 2.80 GHz
($F_{28}$) and the daily RSN since 1947. And after 1965-01-01, the
data set of $F_{28}$ and RSN is continuous in everyday without
even one day drop. As we know $F_{28}$ is most sensitive to the
solar activity (Kundu, 1960) and can be regarded as the most
effective indicator for analyzing the solar activity, it is also
the frequency where the strongest correlations of the radio
emission with the sunspot number and the ionization index of the
coronal E-layer occurred. So, we adopt the daily RSN and $F_{28}$
during from 1965-01-01 to 2011-12-31 to study the mid-term solar
periodicity, the data length is $L=17166$ days. In this work, we
are interested in the long-term periodicity with timescale longer
than 10 yr and the mid-term periodicity with timescale between 27
d (the rotation period of the Sun) and 10 yr. In order to reduce
the sum of calculation, we transform daily data into 5-day
averaged data when we investigate the mid-term periodic modes of
solar activities.

\subsection{Analysis Method}

In order to investigate the periodicity of solar activities, two
methods are adopted to analyze the time series observation data
and present a comparison.

(1) Fast Fourier Transformation

The first one is the fast Fourier transformation (FFT), which can
decompose a mixed time series signal into a series of sine or
cosine components with different periods. An obvious periodic mode
is represented by a sharp peak on the profile of Fourier power
spectra of FFT analysis. The bottom panel of Figure 1 is the power
spectrum of FFT analysis on the annual averaged RSN (ASN) during
1700 - 2011, and this power spectrum can provide the information
of the solar long-term periodic modes. Figure 2 shows the Fourier
power spectra profile of FFT analysis on $F_{28}$ and RSN from
1965-01-01 to 2011-12-31, and these spectrums can provide the
information of solar mid-term periodic modes.

\begin{figure}   
\begin{center}
 \includegraphics[width=8.4 cm]{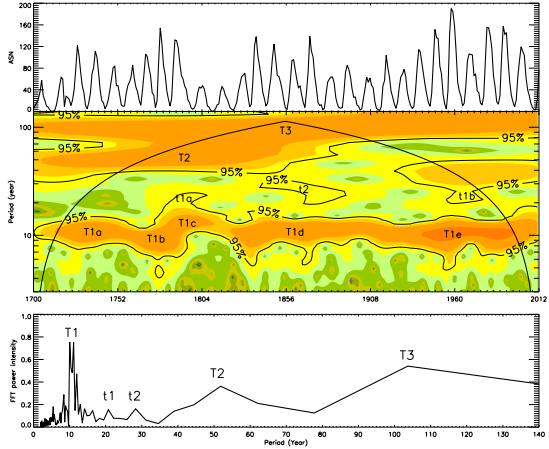}
  \caption{The long periodic modes of solar activity derived from the annual averaged relative sunspot number (ASN)
  during 1700 - 2011. The upper panel is the profile of ASN, the middle panel is the power spectrogram of Morlet wavelet analysis,
  and the lower panel is the corresponding Fourier power spectra of the Fast Fourier transformation.}
\end{center}
\end{figure}

\begin{figure}   
\begin{center}
 \includegraphics[width=8.4 cm]{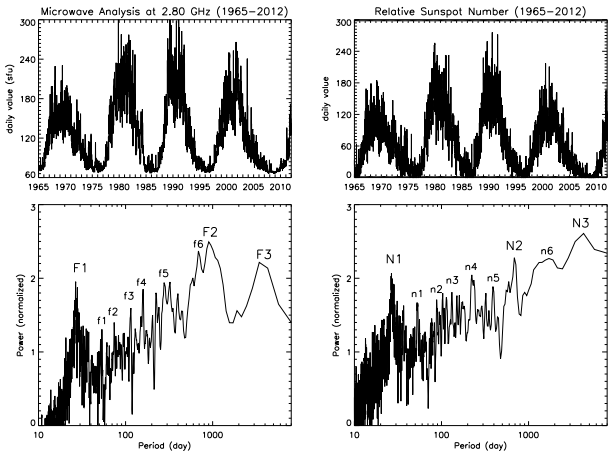}
  \caption{The comparison of the profile (upper panels) and Fourier power spectra of the Fast Fourier Transformation (lower panels) between
  the solar microwave emission flux at frequency of 2.80 GHz and the daily relative sunspot number (RSN) from 1965-01-01 to 2011-12-31.}
\end{center}
\end{figure}

The estimation of frequency errors of FFT analysis can be obtained
by the Nyquist theorem: $\frac{1}{2L}$, where $L$ is the length of
time series of the data. The period obtained from the peak value
of FFT spectrum symbolized as $P_{0}$, the corresponding frequency
is $f_{0}=1/P_{0}$, then the frequency range is $f_{0}\pm 1/(2L)$.
The period range is from $1/(f_{0}+\frac{1}{2L})$ to
$1/(f_{0}-\frac{1}{2L})$, approximately,

\begin{equation}
P\approx P_{0}\pm\frac{P_{0}^{2}}{2L}.
\end{equation}

(2) Wavelet Transformation

The second method is wavelet transform analysis, which is a
powerful tool for analyzing localized power variations in a time
series data set, and can give the scale and time position of a
periodic phenomenon occurred. Here we use the Morlet wavelet
transformation (MWT) to determine the possible periodic modes
concealed in the long-term observation data of ASN, RSN, and
$F_{28}$. MWT is defined as a complex sine wave localized in a
Gaussian window (Morlet et al. 1982). The result of MWT can be
plotted as a spectrogram in time-period space, where an obvious
period mode is represented by a bright strip, the brighter the
strip, the higher the confidence level. From the time-period
spectrum, we may find out when the periodic mode occurred, and how
the period changes with time.

The middle panel of Figure 1 is the power spectrogram of MWT on
ASN during 1700-2011, which presents the detailed information of
solar long-term periodic modes.

Generally, the main features of solar activity are extremely
different from the peak years to the vale years in solar Schwabe
cycles. The daily $F_{28}$ and RSN covers more than 4 solar
Schwabe cycles and 17166 d, the whole power spectrum of MWT
indicates that the periodic modes around the peak years of the
solar cycles are different from that around the vale years of the
solar cycles. So, when we investigate the solar mid-term periodic
modes we divide the 5-day averaged data set of each solar cycle
into peak years and vale years. The peak years of cycle 20-23 are
1966-1972, 1978-1984, 1988-1994, and 1999-2005; the vale years of
cycle 20-23 are 1972-1978, 1984-1988, 1994-1999, and 2005-2011.
Then we make MWT analysis on each data paragraph, respectively.
Figure 3 and Figure 4 are the MWT power spectrograms on the 5-day
averaged F28 and the 5-day averaged RSN around peak years and vale
years in solar Schwabe cycle 20-23, respectively, which may
present the detailed information of solar mid-term periodic modes.

\begin{figure}   
 \includegraphics[width=8.4 cm]{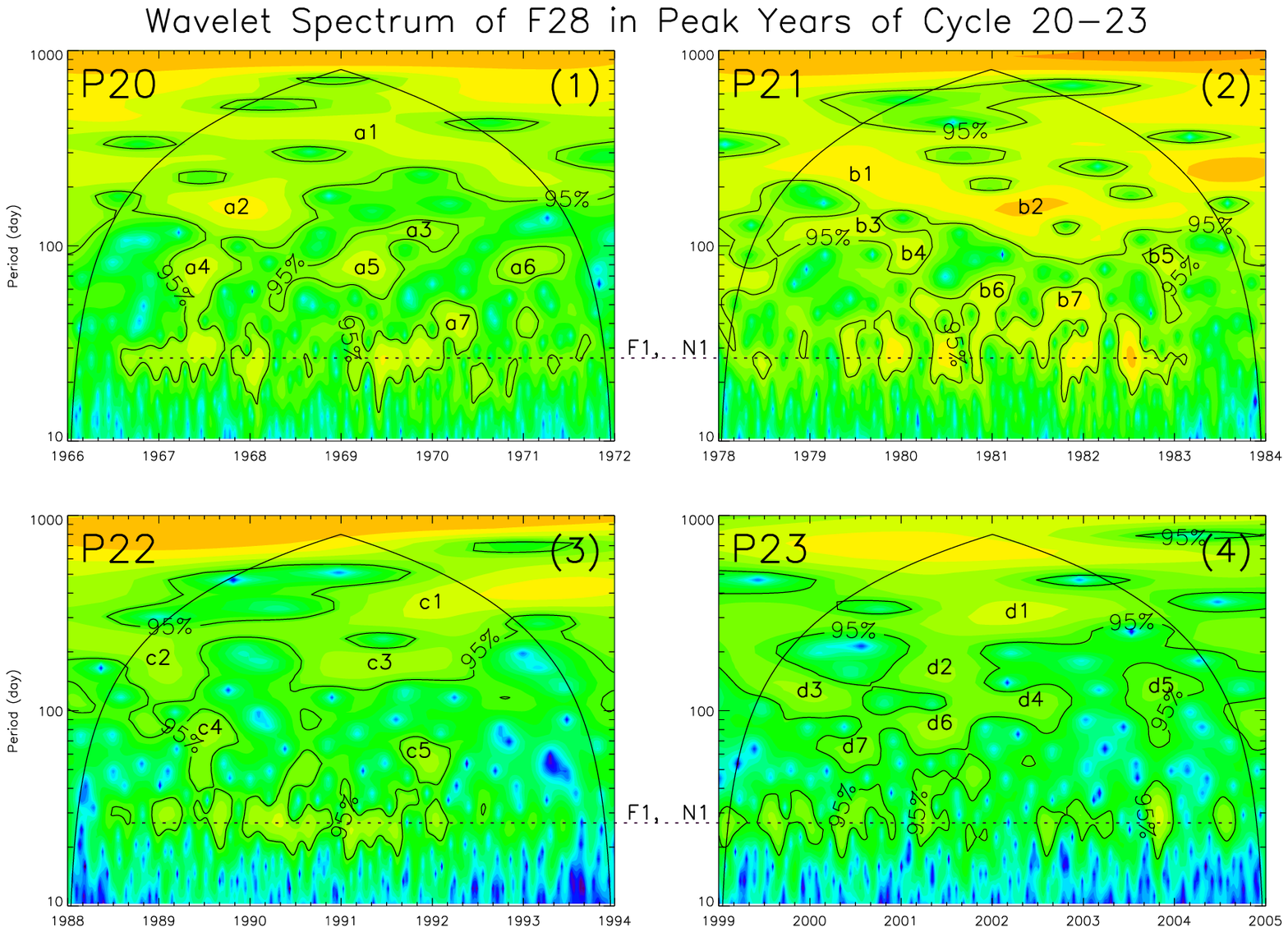}
 \includegraphics[width=8.4 cm]{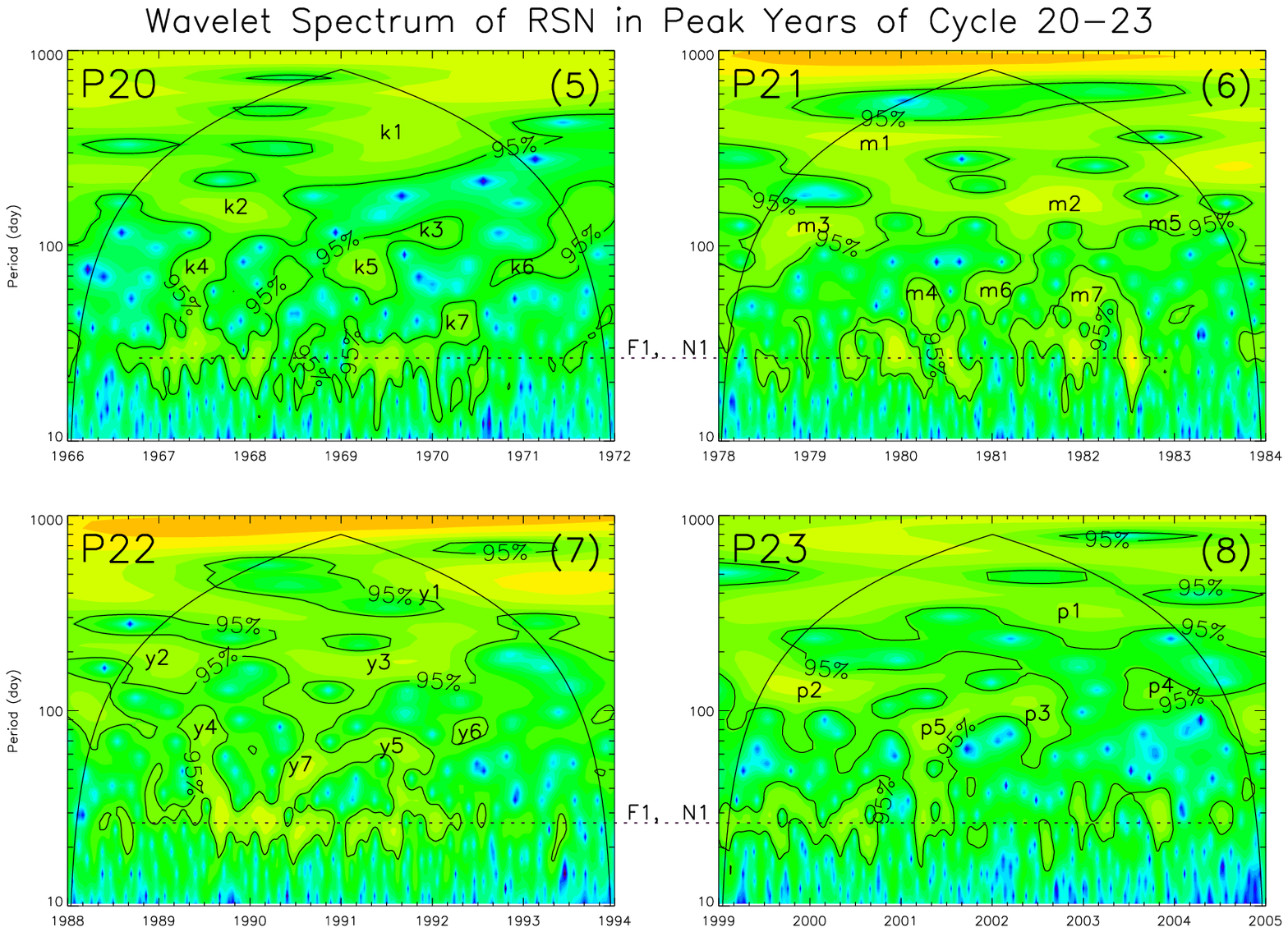}
  \caption{The comparison of the power spectrogram of MWT analysis between the 5-day averaged F28 and the 5-day averaged RSN
  around the peak years in solar Schwabe cycle 20-23.}
\end{figure}

\begin{figure}   
 \includegraphics[width=8.4 cm]{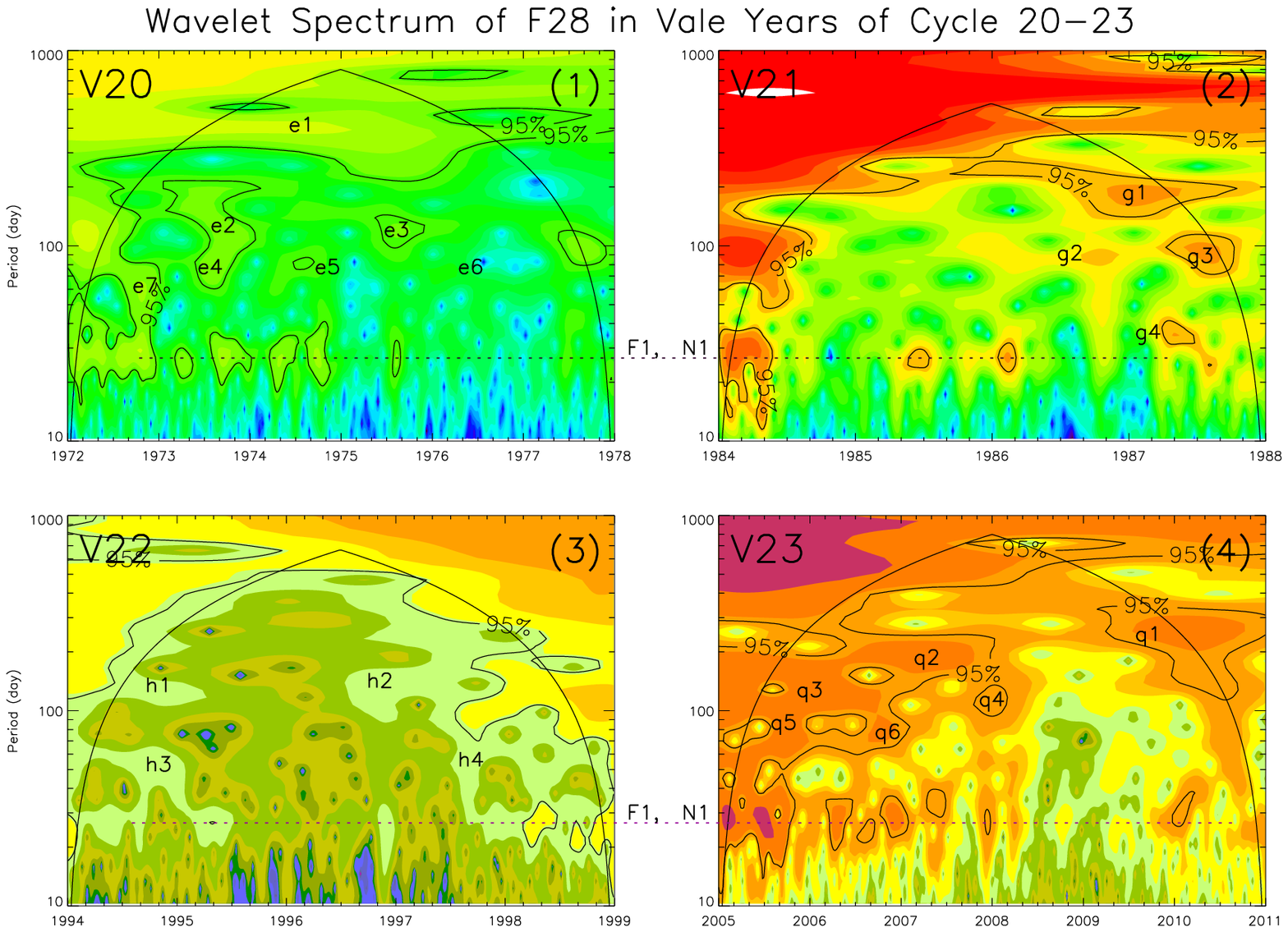}
 \includegraphics[width=8.4 cm]{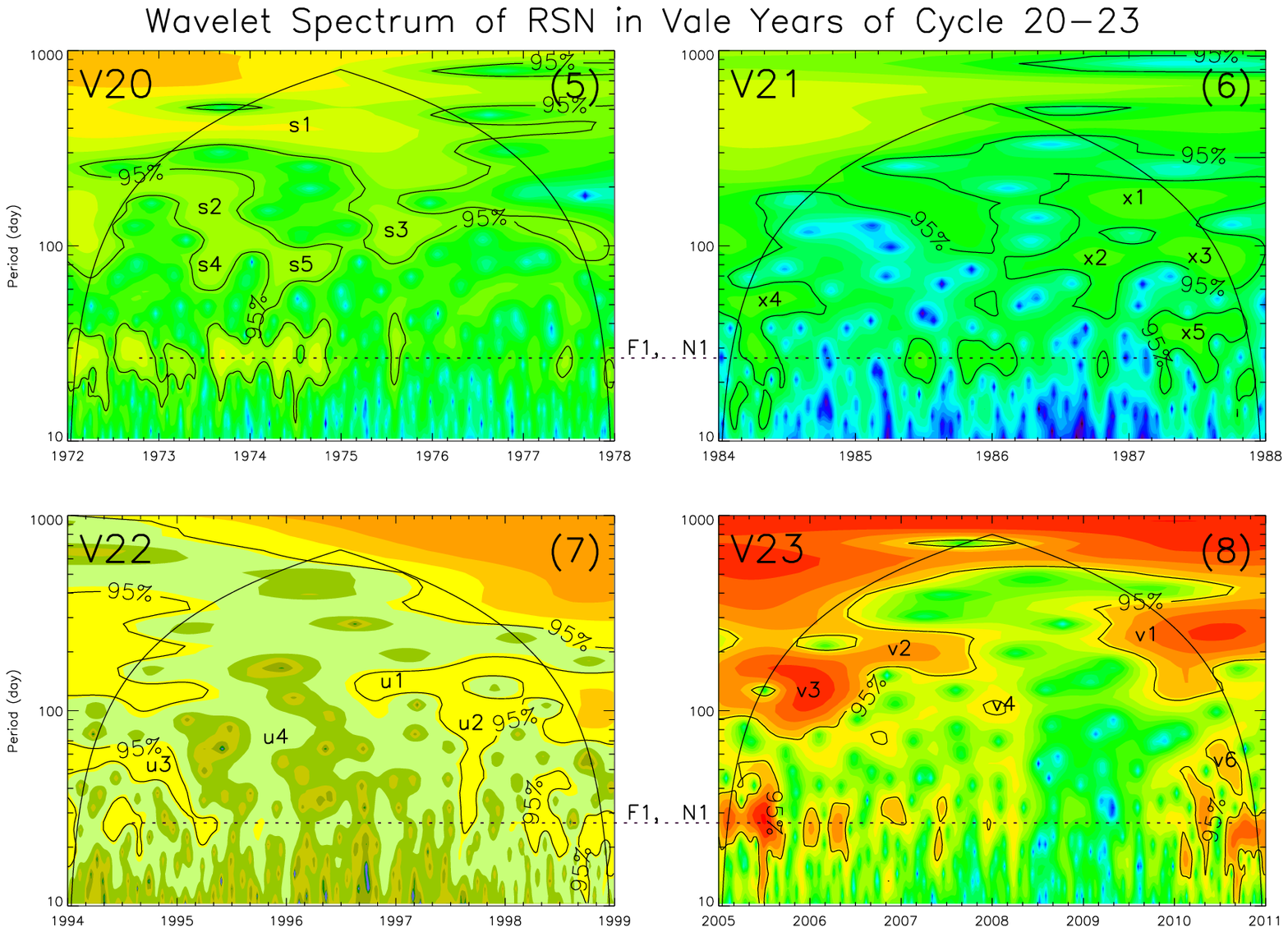}
  \caption{The comparison of the power spectrograms of MWT analysis between the 5-day averaged F28 and the 5-day averaged RSN
  around the vale years in solar Schwabe cycle 20-23.}
\end{figure}

In each panel, we find that there is an identical periodic mode
marked by the horizontal dotted line, and the corresponding period
is about 27 d. This mode is just the one derived from the above
FFT power spectra (F1 and N1). Obviously, it is most possibly
generated from the solar rotation. We do not discuss it in this
work.

\section{Main Results of the Solar Periodic Modes}

\subsection{The Long-term Periodicity}

The first glance at the profile of ASN in the upper panel of
Figure 1 may tell us that there are at least two distinct
different periods: one has a timescale of about 11 yr (in the
range of 9.5-13.5 yr), the other has a timescale of about 103 yr.
The former is the well-known solar Schwabe cycle. The later is
always called as solar secular cycle (Gleissberg, 1939, etc.).

The Fourier power spectra of FFT analysis in the lower panel of
Figure 1 indicates that there are at least several periodic
components: three periods with the confidence level of 95\%: 11 yr
(marked as T1), $52\pm4.3$ yr (T2, 47.7-56.3 yr), and $103\pm17$
yr (T3, 86-120 yr); two periods with the confidence level of 90\%:
$21.5\pm0.74$ yr (t1, 20.76-22.24 yr) and $28.5\pm1.3$ yr (t2,
27.2-29.8 yr). Here, the errors are obtained from Equation (1). In
fact, from a scrutinizing of the FFT spectrum, we may find that
the famous Schwabe Cycle (T1) is a mixture of 3 different
components: $10\pm0.16$ yr, $11.1\pm0.20$ yr, and $11.9\pm0.23$
yr. t1 is always regarded as the solar magnetic cycle with 22 yr
period.

The power spectrogram of MWT analysis on ASN in the middle panel of
Figure 1 shows that the Schwabe cycle is a bright band around 11 yr
with the band width of 9.5-13.5 yr (T1a-T1e); T2 is relative weak in
the range of 50-55 yr; and T3 is a strong periodic component in the
range of 90-110 yr; t1 (t1a and t1b) and t2 also falls into the 95\%
confidence level on the power spectrum of MWT. The similar results
can also be seen in previous works, for example, the co-existence of
about 11-yr, 52-yr and 103-yr in Le and Wang (2003), Tan (2011).
Otaola and Zenteno (1983) reported the evidence of 21.5-yr and
28.5-yr solar cycles.

\subsection{The Mid-term Periodicity}

From the investigation of 5-day averaged $F_{28}$ and RSN from
1965-01-01 to 2011-12-31, we obtain the details of the mid-term
periodicity of solar activity. At first, the FFT power spectra of
5-day averaged $F_{28}$ shows the following mid-term periodic modes:
$900\pm23.6$ d (marked as F2), $53\pm0.08$ d (f1), $75\pm0.16$ d
(f2), $115\pm0.39$ d (f3), $145\pm0.61$ d (f4), $280\pm2.3$ d,
$230\pm1.5$ d, $330\pm3.2$ d and $400\pm4.7$ d (f5 and its
adjacent), and $680\pm13.5$ d (f6) (left panels of Figure 2). The
FFT power spectra of 5-day averaged RSN can show the evidence of
mid-term periodic modes with confidence level $>95\%$: $690\pm13.9$
d (N2), $54\pm0.16$ d (n1), $90\pm0.24$ d, $105\pm0.32$ d, and
$116\pm0.39$ d (n2 and its adjacent), $140\pm0.57$ d (n3),
$235\pm1.6$ d (n4), $395\pm4.5$ d (n5) (right panels of Figure 2).

The panels (1), (2), (3), and (4) of Figure 3 are the MWT power
spectrum of 5-day averaged F28 around the peak years of solar
cycle 20-23, respectively. Several mid-term periodic modes are
noted. In cycle 20 (panel 1): 360 d (a1), 145 d (a2), 115 d (a3),
85 d (a4 and a5); in cycle 21 (panel 2): 235 d (b1), 148 d (b2),
120 d (b3), 90 d (b4 and b5), 54 d (b6), 50 d (b7); in cycle 22
(panel 3): 370 d (c1), 200 d (c2); 180 d (c3), 88 d (c4), and 60 d
(c5); in cycle 23 (panel 4): 320 d (d1), 150 d (d2), 118 d (d3 and
d5), 115 d (d4), 88 d (d6); and 60 d (d7).

The panels (5), (6), (7), and (8) of Figure 3 are the MWT power
spectrum of 5-day averaged RSN around the peak years of solar
cycle 20-23, respectively. Several mid-term periodic modes can be
noted. In cycle 20 (panel 5): 370 d (k1), 145 d (k2), 116 d (k3);
80 d (k4, k5, and k6); in cycle 21 (panel 6): 360 d (m1), 150 d
(m2), 120 d (m3 and m5), 54 d (m4, m6, and m7); in cycle 22 (panel
7): 400 d (y1), 190 d (y2), 180 d (y3), 85 d (y4), 70 d (y5), 83 d
(y6), and 54 d (y7); in cycle 23 (panel 8): 310 d (p1), 120 d (p2
and p4), 90 d (p3), and 83 d (p5).

The panels (1), (2), (3), and (4) of Figure 4 are the MWT power
spectrogram of 5-day averaged F28 around the vale years of solar
cycle 20-23, respectively. The mid-term periodic modes are as
following. In cycle 20 (panel 1): 400 d (e1), 118 d (e2), 115 d
(e3), 80 d (e4 and e6); 86 d (e5), and 60 d (e7); in cycle 21
(panel 2): 300 d (g1), 88 d (g2), and 90 d (g3); in cycle 22
(panel 3): 120 d (h1 and h2), 54 d (h3 and h4); in cycle 23 (panel
4): 240 d (q1), 190 d (q2), 118 d (q3), 100 d (q4), 85 d (q5) and
80 d (q6).

The panels (5), (6), (7), and (8) of Figure 4 are the MWT power
spectrogram of 5-day averaged RSN around the vale years of solar
cycle 20-23, respectively. Several mid-term periodic modes can be
noted: in cycle 20 (panel 5): 400 d (s1), 147 d (s2), 116 d (s3),
85 d (s4), and 86 d (s5); in cycle 21 (panel 6): 180 d (x1), 90 d
(x2 and x3), 54 d (x2 and x3); in cycle 22 (panel 7): 120 d (u1),
85 d (u2), 54 d (u3), 73 d (u4); in cycle 23 (panel 8), 240 d
(v1), 210 d (v2), 118 d (v3), 100 d (v4), and, 54 d (v5).

In brief, the main mid-term periodic modes of solar activity can be
listed as following:

P1: 53-54 d, occurred in Figure 2 (f1 and n1), Figure 3 (b6 and
m4), Figure 4 (h3, h4, x4, x5, and v5), the averaged value is
about 53 d.

P2: 85-90 d, occurred in Figure 2 (n2 adjacent), Figure 3 (a4, a5,
b4, b5, c4, d4, y4, and p3), and Figure 4 (e5, g2, g3, q5, s4, s5,
x2, x3, and u2), the averaged value is about 89 d.

P3: 115-120 d, occurred in Figure 2 (f3 and n2 adjacent), Figure 3
(a3, b3, d3, d5, d4, k3, m3, m4, p2 and p4), and Figure 4 (e2, e3,
h1, h2, q3, s3, u1, and v3), the averaged value is about 117 d.

P4: 140-150 d, occurred in Figure 2 (f4 and n3), Figure 3 (a2, b2,
d2, k2, and m2), and Figure 4 (s2), the averaged value is 146 d.

P5: 230-240 d, occurred in Figure 2 (f5 adjacent, n4), Figure 3
(b1), and Figure 4 (q1 and v1), the averaged value is about 236 d.

P6: 360-370 d, occurred in Figure 3 (a1, c1, and k1), the averaged
value is about 367 d.

P7: 395-400 d, occurred in Figure 2 (f5 adjacent and n5), Figure 3
(y1), and Figure 4 (e1 and s1), the averaged value is about 399 d.

Here, we neglect other periodic modes which are also occurred but
not frequently in the above analysis, such as 60 d, 75 d, 200 d,
320 d, etc. In the previous literatures, several similar mid-term
periodic modes of solar activity are also reported. Kilcik et al.
(2010) show the existence of 53 d, 85 d, 115 d, 137 d, 152 d, and
248 d periodic modes derived from solar flare index. Pap, Tobiska,
and Bouwer (1990) gave the evidence of 113 d and 237 d periodic
modes. Caballero \& Valdes-Galicia (2003) reported the 89 d and
115 d periodic modes from the records in a high altitude neutron
monitor. The famous Rieger period with 138-159 d is frequently
reported from publications (Rieger, et al. 1984, etc.).

\subsection{Comparison between the solar cycles and the planetary motions}

It is well known that the most obvious periodic modes in solar
system are the planetary motions around the Sun. These periodic
motions include planetary sidereal motions around the Sun
producing tidal forces and the conjunction motions of two or more
planets producing spring tidal forces on solar mass, and the
latter can be called as spring tidal period. The spring tidal
period of two planets can be calculated by:
$P=\frac{1}{2}\frac{P1\cdot P2}{P1-P2}$, and its integral
multiples are also spring tidal periods. Here $P1$ and $P2$ are
the sidereal periods of the two planets, respectively. For
example, we may obtain the spring tidal period of Mercury and
Jupiter at 89.77 d, and 115.89 d for Mercury and Earth, 144.55 d
for Mercury and Venus, 236.97 d for Venus and Jupiter, 398.88 d
for Earth and Jupiter, etc. It has been found that Venus, Earth
and Jupiter tend to be mostly aligned every 11.07 yr (Takahashi,
1968; Scafetta, 2012a, etc.). With these periodic motions, the
planetary gravitational forces acting on solar mass will undergo
periodic variations. It is questionable whether these periodic
variations of the planetary forces can trigger the periodic solar
activity or not. However, it is also meaningful to make a
comprehensive comparison between the solar periodic modes and the
planetary motions.

Here, we define a parameter $\beta$ to measure the similarity
between the solar activity periodic modes and the corresponding
planetary periods:

\begin{equation}
\beta=\frac{|P_{sa}-P_{sp}|}{P_{sp}}.
\end{equation}

Here $P_{sa}$ is the period of solar activity, and $P_{sp}$ is the
period of planetary sidereal motions or the conjunction motion of
two or more planets. We find that there are 9 periodic modes of
solar activity to be very close to some planetary sidereal periods
or spring tidal periods of two or more planets. We may name such
kind of periodic modes as planetary-like cycles (PLC), and
symbolize them as:

PLC1: P2 (85-90 d, averaged value 89 d) is close to the spring
tidal period of Mercury and Jupiter (89.77 d), the averaged $\beta
\approx 0.87\%$. The Mercury sidereal period (88.0 d) is also very
close to P2, and the value of $\beta$ is about 1.13\%.

PLC2: P3 (115-120 d, averaged value 117 d) is close to the spring
tidal period of Mercury and Earth (115.89 d), the averaged $\beta
\approx 0.95\%$.

PLC3: P4 (140-150 d, averaged value 146 d) is close to the spring
tidal period of the Mercury and Venus (144.55 d), the averaged
$\beta \approx 0.99\%$.

PLC4: P5 (230-240 d, averaged value 236 d) is close to the spring
tidal period of Venus and Jupiter (236.97 d), the averaged $\beta
\approx 0.41\%$.

PLC5: P6 (360-370 d, averaged value 367 d) is close to the
sidereal period of Earth (365.25 d), $\beta \approx 0.48\%$.

PLC6: P7 ($\sim$400 d) is close to the spring tidal period of
Earth and Jupiter (398.88 d), the averaged $\beta \approx 0.28\%$.

PLC7: the overall effect of T1 (9.5-13 yr with the averaged value
11.2 yr) is the most prominent Schwabe 11-yr cycle which is always
regarded as having some links with the Jupiter's motion. In fact,
PLC7 is actually a mixture of 3 different components: PLC7-1:
$10\pm0.16$ yr (9.84-10.16), very close to a half of the spring
tidal period of Jupiter and Saturn (9.93 yr). The averaged values
of $\beta$ is 0.70\%; PLC7-2: $11.1\pm0.20$ yr (10.90-11.30), very
close to period of the conjunction motion of Venus, Earth and
Jupiter (11.07 yr). The averaged values of $\beta$ is 0.27\%; and
PLC7-3: $11.9\pm0.23$ yr (11.67-12.13), very close to Jupiter
sidereal period (11.86 yr). The averaged values of $\beta$ is
0.31\%.

PLC8: t1 ($21.5\pm0.74$ yr) is close to the synodic period of
Jupiter and Saturn (19.87 yr), and the averaged value of $\beta$
is 8.2\%. There is an another possibility that the 21.5-yr
periodic mode is a resonance mode of 2 times of the 11.07-yr
conjunction motion period of Venus, Earth and Jupiter (Scafetta,
2012b). In such regime, $\beta\sim 3\%$. Furthermore, as the error
range also beyond 3\%, so far, we have no enough evidences to
determine which one is true.

PLC9: t2 ($28.5\pm1.3$ yr) is close to the sidereal period of
Saturn (29.42 yr), the difference is from -2.22 yr to 0.38 yr, and
the averaged value of $\beta$ is 3.1\%.

The above results indicate that there are only a few not many
periodic modes of solar activities seemed to have no relationship
with the planetary motions. Most of the periodic modes of solar
activities are PLCs, possibly associated to the planetary motions.
According to the values of $\beta$, we may classify these PLCs
into two kinds:

(1) Strong PLC, which has relatively short period ($P<12$ yr), low
$\beta$ value, $\beta<1.0\%$, and related to the motions of the
inner planets and the Jupiter, including PLC1, PLC2, PLC3, PLC4,
PLC5, PLC6, and PLC7 (at the same time, PLC7 contains 3 different
periodic modes).

(2) weak PLC, which has relatively long period ($P>12$ yr), high
$\beta$ value, $\beta>3.0\%$, and possibly related to the motions of
outer planets, including PLC8 and PLC9. At the same time, we also
notice that the intrinsic error exceeds 3\% when the period beyond
20 yr. Therefore, $\beta<1.0\%$ is only a moderate parameter when we
determine which mode belongs to weak PLC.

There are also several periodic modes which seems no obvious
relationship with planetary motions. We may call them as Non-PLC.
This class includes the modes with period of 53-54 d, 70-75 d,
$52\pm4.3$ yr (T2, 47.7-56.3 yr), and $103\pm17$ yr (T3, 86-120
yr).

\section{Physical Discussions}

The relatively close equality of the averaged solar Schwabe cycle
period and the sidereal period of Jupiter was noticed for a long
time. Since the 19th century a theory has been proposed claiming
that the solar activity is partially driven by the planetary tidal
forces (Wolf, 1859; Jose, 1965; Wood, 1975; Landscheidt 1999).
However, most existing literature focuses on discussing the
relationship between the solar Schwabe cycle with period of about
11 yr and the orbital motion of Jupiter. The physics of the
periodic modes of solar activity with multiple timescales is still
an open question. We need to answer the following questions: how
strong the planetary gravitational force acting on the solar
plasmas? what kind of planetary motion dominate which solar
periodic mode? what is the physical mechanism of the PLC?

Many people believe that $F_{td}$ acting on the solar mass may
result in the solar activity (Jose 1965, Landscheidt 1999,
Scafetta 2012). As the tachocline layer is most possibly the
source region of the solar magnetic fields (Parfrey and Menou
2007, etc.), similar to the other authors, we also calculate
$F_{td}$ at tachocline layer:

\begin{equation}
F_{td}=\frac{2GM_{p}R_{t}}{a_{p}^{3}}=\frac{GM_{p}}{a_{p}^{2}}\frac{2R_{t}}{a_{p}}
\end{equation}

Here, $M_{p}$ is the planet mass, $R_{t}$ is the solar radius at
tachocline level, $a_{p}$ is the averaged distance between the Sun
and the planet. Here we define a force unit $sgu$ as 1\%
gravitational force acting on unit solar mass operated by Earth,
1$sgu=1.782\times 10^{-10} m\cdot s^{-2}$. During the spring
tides, $F_{td}$ is the sum of the tidal forces of the
corresponding planets. Table 1 lists a series of tidal forces,
including several spring tidal forces.

Furthermore, Equation (3) indicates that the tidal force is
proportional to the radius. The solar photospheric mass will
suffer a stronger tidal force than that of the solar internal
mass. Even if the solar magnetic field possibly originates at the
tachocline level, solar eruptions (solar flares, CME, and eruptive
filaments, etc.) take place in the solar atmosphere, especially in
the solar chromosphere and corona.

\begin{table}\def~{\hphantom{0}}
  \begin{center}
  \caption{Planetary tidal forces (unit in $sgu$), listed by the decreasing order. $P_{td}$ is the corresponding period.
  Here abbreviations of Mc, V, E, Mr, J, S, U, and Np express Mercury, Venus, Earth, Mars, Jupiter, Saturn, Uranus, and
  Neptune, respectively. The unit of $\phi$ is $sgu/yr$.}
  \label{tab:kd}
  \begin{tabular}{lcccccccc}\hline
  planet &  $F_{td}$ & $P_{td}$ &  $\phi$    & $P_{sa}$ & $\beta(\%)$ \\\hline
  V/E/J  &   3.53    & 11.07yr &   0.637    &  11.1yr   &  0.27       \\
  V/J    &   2.88    &  236.9d &   8.865    &  236d     &  0.41       \\
  E/J    &   2.12    &  398.9d &   3.888    &  400d     &  0.28       \\
  Mc/J   &   2.09    &   89.8d &   17.02    &  89d      &  0.87       \\
  Mc/V   &   2.03    &  144.6d &   10.23    &  146d     &  0.99       \\
  J/S    &   1.54    &  9.93yr &   0.310    &  10.0yr   &  0.70       \\
  J/S    &   1.54    & 19.87yr &   0.156    &  21.5yr   &  8.20       \\
  J      &   1.47    & 11.86yr &   0.248    &  11.9yr   &  0.31       \\
  V      &   1.40    &  224.7d &   4.564    &   --      &   --        \\
  Mc/E   &   1.27    &  115.9d &   8.023    &  117d     &  0.95       \\
  E      &   0.65    &  365.2d &   1.304    &  367d     &  0.48       \\
  Mc     &   0.62    &   88.0d &   5.155    &   --      &   --        \\
  S      &   0.07    & 29.42yr &   0.004    &  28.5yr   &  3.13       \\
  Mr     &   0.02    &  686.9d &   0.021    &           &             \\
  U      &  $<$0.01  & 83.75yr &   $<$0.001 &           &             \\
  Np     &  $<$0.01  & 163.7yr &   $<$0.001 &           &             \\\hline
  \end{tabular}
 \end{center}
\end{table}

De Jager and Versteegh (2005) compared the three accelerations
working on the solar matter at the tachocline level: the
acceleration due to $F_{td}$, the one due to the motion of the sun
around the mass center of the solar system with the sum of
planetary attractions, and the actual one due to connective
motions in the tachocline level and above it. Supposed an averaged
solar convective velocity at the bottom of the convection zone,
then the actual acceleration near the tachocline level is about
$6\times10^{-6} m\cdot s^{-2}\simeq$ $3.36\times10^{4}$ sgu
(Robinson et al. 2004). They found that the later is much larger
than the former two by several orders of magnitude, and pointed
out that the planetary attraction is not the cause of solar
activity, the main cause is purely originated from the solar
interior. Callebaut, de Jager and Duhau (2012) reconsidered the
internal convective velocity, examined the magnetic buoyancy and
Coriolis force, and deduced that the planetary influences are too
weak to be even a small modulation of the solar cycles.

However, as we know that solar atmosphere is a huge plasma system,
any equilibrium in such system is always flimsy and most unstable,
even with a very small perturbation, the system will lost its
equilibrium, and a variety of instabilities are very easy to
develop, accumulate, and trigger a considerable outburst.
Recently, Shravan, Thomas, and Katepalli (2012) use techniques of
helioseismology to image the flows in solar interior, and find
that the solar convective velocities are 20-100 times weaker than
the previous theoretical estimations. If this fact is true, it
will reduce the gap between the solar interior convective velocity
and the planetary tidal motions to some extent.

In order to investigate the change rate of planetary tidal force
acting on solar mass operated by the planets during each PLC, we
define another parameter, $\phi$, it can be calculated by:

\begin{equation}
\phi=\frac{F_{td}}{(1/2)P_{sp}}.
\end{equation}

Here, $F_{td}$ is the planetary tidal force or the spring tidal
force of conjunction motions of two or more planets, the unit of
$\phi$ is sgu/yr. The $\phi$ values are listed in the fourth
column of Table 1. As a comparison, the periods and $\beta$ are
also listed in fifth and sixth column.

We find that the tidal force of conjunction motion of Venus, Earth
and Jupiter is the strongest one (3.53 sgu) with period of 11.07
yr, which is related to the solar periodic mode of 11.1 yr, this
mode is just the one closest to the central period of the most
famous Schwabe Cycle. Table 1 shows that most tidal forces
associated with PLCs have the value $F_{td}>1.0$ sgu. The sidereal
tidal force of Venus is 1.40 sgu, but its period (224.7 d) is very
close to the spring tidal period of Venus and Jupiter
($F_{td}=2.88$ sgu) and easy to be submerged in the latter.

By adopting the parameter $\phi$, we find that all the strong PLCs
have the values $\phi>0.2$ sgu/yr, and $\phi<0.2$ sgu/yr for all
weak PLCs. In this regime, we find that the biggest $\phi$ value
is related to the conjunction motion of the Mercury and Jupiter
(17.02 sgu/yr), the second biggest $\phi$ value is related to the
conjunction motion of Mercury and Venus (10.23 sgu/yr). The $\phi$
values of the sidereal motions of Mercury is very big, however,
because the Mercury sidereal period is very close to the spring
tidal period of Mercury and Jupiter, which is very easy to be
submerged in the latter. The Venus's condition is similar to the
Mercury's.

Table 1 indicates that tidal forces operated by Mars, Uranus, and
Neptune are very weak relative to the other 5 planets by several
orders of magnitudes, that we may neglect their tidal effects on
solar activities. Jupiter plays a key role in most periodic modes
by its own sidereal motion or conjunction motion with other
planets (PLC1, PLC4, PLC5, PLC7-1, PLC7-2, PLC7-3, and PLC8).
Additionally, some of inner planets, such as Mercury, Venus and
Earth also play important roles in the formation of PLCs.

The conjunction motion and the related spring tidal forces of the
planets dominate the formation of most PLCs. Such as the
conjunction motion of Venus/Earth/Jupiter, Venus/Jupiter,
Earth/Jupiter, Mercury/Jupiter, Mercury/Venus, Mercury/Earth, and
Jupiter/saturn. This is understandable, because the strongest
tidal force acted by any individual planet is only 1.47 sgu
(Jupiter), while many spring tidal forces acted by conjunction
motion of two or more planets are stronger than that acted by
Jupiter alone. As mentioned above, the conjunction motion of
Venus, Earth and Jupiter forms the strongest periodic mode of
11.07 yr and related to the most famous Schwabe cycle.

\begin{figure}   
 \includegraphics[width=8.4 cm]{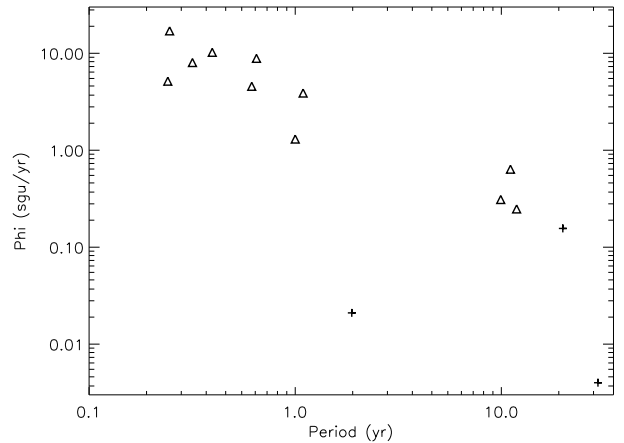}
  \caption{The relationships between period (P) and the change rate of planetary tidal force
acting on solar mass operated by the planets (Phi). Triangle
represents the strong PLC, and plus sign represents the weak PLC.}
\end{figure}

Figure 5 shows the relationships between period ($P$) and $\phi$.
It indicates that PLCs have an obvious tendency of the value
$\phi$ decreasing with the period increasing. Particularly, strong
PLCs (triangle signs) are much more obvious with such tendency,
while the weak PLCs (plus signs) are more disperse.

Some people will also suspect that $F_{td}$ is still too weak to
excite a considerable perturbation in solar atmosphere. And the
difference between the planetary periods and the periods of PLC
also let us to distrust the relationships between planetary motion
and the solar periodical activity. As we know that solar
atmosphere is a huge magnetized plasma system, any equilibrium in
such system is flimsy and most unstable. Even if a very small
perturbation can also disrupt the magnetized plasma system, make
it lost the equilibrium, and trigger a considerable instability.
Such instability can accumulate in a complicated way, and result
in energy releasing. Finally, the solar activity bursts out. The
enormous energy released in the solar activity is essentially
originated from the solar inner processes. Here, $F_{td}$ just
plays a role of blasting fuse of solar activity. Recently,
Scafetta (2012b) proposed that the planetary tidal gravitational
energy can amplify the nuclear fusion in the solar core region,
and result in the solar periodic activities. Maybe this is a
possible physical mechanism. Based on the tidal cycles of the
Jupiter and Saturn plus the dynamo theory, Scafetta (2012a) has
shown that it is possible to reconstruct solar variability with a
high accuracy at the decadal, secular and millennial timescale
throughout the Holocene.

\section{Conclusions}

This work presents a comprehensive investigations of the
quasi-periodic modes of solar activity with multi-timescale and
their relationships with the planetary periodic motions. From this
investigation, we may get the following conclusions:

(1) All the periodic modes of solar activities can be classified
into 3 classes:

A. Strong PLC, which is related strongly with planetary motions.
This class includes 8 periodic modes: PLC1 (85-90 d, averaged
period is 89 d, close to the spring tidal period of Mercury and
Jupiter), PLC2 (115-120 d, averaged period is 117 d, close to the
spring tidal period of Mercury and Earth), PLC3 (140-150 d,
averaged period is 146 d, close to the spring tidal period of
Mercury and Venus), PLC4 (230-240 d, averaged period is 236 d,
close to the spring tidal period of Venus and Jupiter), PLC5 (400
d, close to the spring tidal period of Earth and Jupiter), PLC6
(360-370 d, close to the sidereal period of Earth), PLC7-1 (10 yr,
close to the spring tidal period of Jupiter and Saturn), PLC7-2
(11.1 yr, close to the spring tidal period of Venus, Earth and
Jupiter), PLC7-3 (11.9 yr, close to the sidereal period of
Jupiter). This class mode has relatively short period ($P<12$ yr),
$\phi>0.2$ sgu/yr, low $\beta$ value, $\beta<1.0\%$, and related
to the motions of the inner planets and of Jupiter.

B. Weak PLC, which is related weakly with planetary motions. This
class includes 2 periodic modes: PLC8 (21.5 yr, close to the
synodic period of Jupiter and Saturn) and PLC9 (28.5 yr, close to
the sidereal period of Saturn). This class mode has relatively
long period ($P>12$ yr), $\phi<0.2$ sgu/yr, high $\beta$ value,
$\beta>3.0\%$, and possibly related to the motions of outer
planets. At the same time, because periods of weak PLCs are always
very long, and their error levels are always very large, the
uncertainty of weak PLCs can not be neglected.

C. Non-PLC,  which is so far regarded as no obvious relationship
with any planetary motions. This class includes the modes with
period of 53-54 d, 70-75 d, $52\pm4.3$ yr (T2, 47.7-56.3 yr), and
$103\pm17$ yr (T3, 86-120 yr).

(2) Among all the planets, Jupiter plays a key role in most
periodic modes by its own sidereal motion or conjunction motion
with other planets. 7 PLCs (PLC1, PLC4, PLC5, PLC7-1, PLC7-2,
PLC7-3, and PLC8) are dominated by Jupiter.

(3) Among all the planetary motions, the conjunction motion of the
inner planets dominated the formation of most PLCs. Such as the
conjunction motion of Venus/Earth/Jupiter, Venus/Jupiter,
Earth/Jupiter, Mercury/Jupiter, Mercury/Venus, Mercury/Earth, and
Jupiter/saturn, etc.

The closely connection between the mid-term periodic modes of
solar activity and the planetary motions may imply that it can be
applied to predict the solar activity in the near future. In our
next step researches, we will focus on investigating the effects
of the planetary motions to the solar short- and mid-term
activity, and the detailed processes of the physical mechanism.

\acknowledgments

The authors would like to thank the referee's valuable comments on
the manuscript and SGD teams for the systematic data. This work is
supported by MOST Grant No. 2011CB811401, NSFC Grant No. 11273030,
10921303, and the National Major Scientific Equipment R\&D Project
ZDYZ2009-3.


\begin{thebibliography}{}

\bibitem[Caballero(2003)]{Caballero2003}Caballero, R., \& Valdes-Galicia, J.F.: 2003, \emph{Solar Phys.} \textbf{213}, 413.

\bibitem[Callebaut(2012)]{Callebaut2012} Callebaut, D.K., de Jager, C., \& Huhau, S.: 2012, \emph{J. Atmospher. Sol. Terr. Phys.}, \textbf{80}, 73

\bibitem[de Jager(2005)]{de Jager2005}de Jager, C., \& Versteegh, G.J.M.: 2005, \emph{Solar Phys.} \textbf{229}, 175.

\bibitem[Frick et al(1997)]{Frick97}Frick, P., Galyagin, D., Hoyt, D.V., et al.: 1997, \emph{Astron Astrophys} \textbf{328}, 670.

\bibitem[Gleissberg(1939)]{Gleissberg1939}Gleissberg, W.: 1939, \emph{Observatory} \textbf{62}, 158.

\bibitem[Gleissberg(1971)]{Gleissberg71}Gleissberg, W.: 1971, \emph{Solar Phys.} \textbf{21}, 240.

\bibitem[Hiremath(2009)]{Hiremath2009}Hiremath, K.M.: 2009, arXiv0909.4420[astro-ph.SR].

\bibitem[Jose(1965)]{Jose 1965} Jose, P.D., 1965, \emph{AJ}, \textbf{70}, 193

\bibitem[Kilcik(2010)]{Kilcik2010} Kilcik, A., Ozguc, A., Rozelot, J.P., \& Atas, T. 2010, \emph{Solar Phys.}, \textbf{264},
255.

\bibitem[Kundu(1960)]{Kundu1960} Kundu, M.R., 1960, \emph{J. Grophys. Res.}, \textbf{65},
3903.

\bibitem[Landscheidt(1999)]{Landscheidt 1999} Landscheidt, Th, 1999, \emph{Solar Phys.}, \textbf{189}, 415.

\bibitem[Le(2003)]{Le2003} Le, G.M., \& Wang, J.L.: 2003, \emph{Chin. J. Astron. Astrophys.}, \textbf{3}, 391.

\bibitem[Morlet(1982)] {Morlet 1982} Morlet, J., Arehs, G., Forgeau, I., Giard, D.: 1982, \emph{Geophysics}, \textbf{47}, 203.

\bibitem[Nagovitsyn(1997)]{Nagovitsyn97}Nagovitsyn, Yu A.: 1997, \emph{Astron Lett} \textbf{23}, 742.

\bibitem[Otaola \& Zenteno(1983)]{Otaola03}Otaola, J.A.Q., \& Zenteno, G.: 1983, \emph{Solar Phys.} \textbf{89}, 209.

\bibitem[Parfrey(2007)]{Parfrey2007} Parfrey, K.P., \& Menou, K.: 2007, \emph{ApJ}, \textbf{667}, L207.

\bibitem[Pap(1990)]{pap1990} Pap, J., Tobiska, W.K., \& Bouwer, S.D.: 1990, \emph{Solar Phys.}, \textbf{129}, 165.

\bibitem[Rieger(1984)]{Rieger1984}Rieger, E., Kanbach, G., Reppin, C., Share, G.H., Forrest, D.J., \& Chupp, E.L.: 1984, \emph{Nature} \textbf{312}, 623.

\bibitem[Robinson(2004)] {Robinson 2004} Robinson, F.J., Demarque, P, Li, L.A., Sofia, S., Kim, Y.-C., Chan, K.L., Guenther, D.B.: 2004,
\emph{Mon. Not. R. Astron. Soc.}, \textbf{340}, 923.

\bibitem[Scafetta1(2012a)]{Scafetta2012a} Scafetta, N.: 2012a, \emph{J. Atmospher. Sol. Terr. Phys.}, \textbf{80}, 296.

\bibitem[Scafetta2(2012b)]{Scafetta2012b} Scafetta, N.: 2012b, \emph{J. Atmospher. Sol. Terr. Phys.}, \textbf{81-82}, 27.

\bibitem[Schove(1979)]{Schove79}Schove, D.J.: 1979, \emph{Solar Phys.} \textbf{63}, 423.

\bibitem[Shravan(2012)]{Shravan2012}Shravan, M.H., Thomas, L.D., Jr., Katepalli, R.S.: 2012, \emph{PNAS} \textbf{109}, 11928.

\bibitem[Suess(1980)]{Suess80}Suess, H.E.: 1980, \emph{Radiocarbon} \textbf{20}, 200.

\bibitem[Takahashi(1968)]{Takahashi1968} Takahashi, K.: 1968, \emph{Solar Phys.}, \textbf{3}, 598.

\bibitem[Tan(2011)]{Tan2011} Tan, B.L. 2011, \emph{Ap$\&$SS}, \textbf{332}, 65.

\bibitem[Yoshimura(1978)]{Yoshimura1978} Yoshimura, H. 1978, \emph{ApJ}, \textbf{226}, 706.

\bibitem[Wolf(1859)] {Wolf1859} Wolf, R.: 1859, \emph{Mon. Not. R. Astron. Soc.}, \textbf{19}, 85.

\bibitem[Wolff(2010)]{Wolff2010} Wolff, C.L., \& Patrone, P.N.: 2010, \emph{Solar Phys.}, \textbf{266}, 227.

\bibitem[Wood(1975)]{Wood75}Wood, R.M.: 1975, \emph{Nature} \textbf{255}, 312.

\end{thebibliography}

\end{document}